\documentclass{article}
\usepackage{spconf,amsmath,amssymb,mathrsfs}
\usepackage{microtype}
\usepackage{algorithm,algpseudocode}
\usepackage{graphicx,epsfig}
\usepackage[caption=false,font=footnotesize]{subfig}
\def\O{\mathbf{O}}

\def\t{\boldsymbol{t}}
\def\x{\boldsymbol{x}}

\def\a{\boldsymbol{a}}
\def\C{\mathbf{C}}
\def\Z{\mathbf{Z}}
\def\I{\mathbf{I}}
\def\e{\boldsymbol{e}}
\def\q{\boldsymbol{q}}
\def\f{\boldsymbol{f}}

\def\J{\mathbf{J}}
\def\L{\mathbf{L}}
\def\B{\mathbf{B}}
\def\D{\mathscr{D}}
\def\W{\mathbf{W}}
\def\Del{\boldsymbol{\delta}}
\def\G{\mathscr{G}}
\def\Gs{\mathbf{G}^{\star}}
\def\V{\mathscr{V}}
\def\E{\mathscr{E}}
\def\P{\mathscr{P}}
\def\A{\mathscr{A}}
\def\Tr{\mathrm{Trace}}

\title{LARGE-SCALE SENSOR NETWORK LOCALIZATION VIA RIGID SUBNETWORK REGISTRATION}

\name{Kunal N. Chaudhury$^{\star\star}$ \qquad Yuehaw Khoo$^\dagger$  \qquad  Amit Singer$^{\star}$}

\address{$^{\star\star}$Department of Electrical Engineering, Indian Institute of Science, India\\ 
$^\dagger$Department of Physics, Princeton University, USA\\
$^\star$Department of Mathematics and PACM, Princeton University, USA.}

\begin{document}
\ninept

\maketitle

\begin{abstract}
In this paper, we describe an algorithm for sensor network localization (SNL) that proceeds by dividing the whole network into smaller subnetworks, then localizes them in 
parallel using some fast and accurate algorithm, and finally registers the localized subnetworks in a global coordinate system. 
We demonstrate that this divide-and-conquer algorithm can be used to leverage existing high-precision SNL algorithms to large-scale networks, which 
could otherwise only be applied to small-to-medium sized networks. The main contribution of this paper concerns the final registration phase.  
In particular, we consider a least-squares formulation of the registration problem (both with and without anchor constraints) and 
demonstrate how this otherwise non-convex problem can be relaxed into a tractable convex program. 
We provide some preliminary simulation results for large-scale SNL demonstrating that the proposed registration algorithm (together with an accurate localization scheme) offers a good tradeoff 
between run time and accuracy.
\end{abstract}


\section{Introduction}

The computational problem in sensor network localization (SNL) is one of determining the position of \textit{sensors} in two or three dimensions from incomplete and inaccurate inter-sensor distances. 
In some cases, the distances between certain fixed \textit{anchors} (whose positions are known fairly accurately) and some sensors are also provided \cite{Patwari2005,Madria2003}.
While the SNL problem in its full generality is known to be computationally intractable \cite{Yemini1979},
%. In particular, it is known to be NP-hard for the realistic SNL model in which 
%the distance between two sensors is assumed to be known if and only if they are within the \textit{radio range} of each other \cite{Aspnes2006}. 
nevertheless, there has been considerable progress on developing 
algorithms that can efficiently solve the SNL problem (exactly or approximately) under appropriate assumptions on the network connectivity, and which are resilient to noises in 
the distances and the anchor positions. 
Popular methods include classical multidimensional scaling \cite{Cox2001}, belief propagation \cite{Win2009}, non-linear filtering \cite{O-S2009}, and 
geometric methods \cite{So2009,Moura2010}. We refer the readers to \cite{Patwari2005,Madria2003,Mao2007} for a broad survey of SNL algorithms.
%A survey of SNL algorithms is beyond the scope of this paper, and instead we refer the readers to these excellent reviews \cite{Patwari2005,Madria2003,Mao2007}.

The distance constraints in SNL make the problem intrinsically non-convex. In the last few years, some very effective convex relaxations of the SNL problem have been proposed \cite{Biswas2006,Tseng2007,Montanari2011}. 
%The general theme here is that some 
%(non-convex) formulation of the problem is considered, which is then relaxed into a convex program whose solution can be computed in polynomial time . 
%We refer the readers to this review \cite{Zhang2010review} for a comprehensive account of some of these and related convex relaxations in signal processing and %communications.  
Apart from offering remarkable localization accuracy in practice, these algorithms also come with guarantees on exact recovery and stability under appropriate assumptions on the network connectivity \cite{Montanari2011,So2007}. 
%Moreover, one can use the solution obtained from one of these relaxations to initialize non-convex methods, which typically require a good initialization \cite{Biswas2006}.
A drawback of these algorithms is that their computational complexity often scales poorly with the network size. 
For example, the convex relaxations in \cite{Biswas2006,Montanari2011} result in semidefinite programs (SDP) with $O(N^2)$ variables, where $N$ is the number of sensors. 
Due to the high memory requirement and computational cost of standard interior-point SDP solvers \cite{SDPT3,SeDuMi}, this limits the scope of these SDP-based methods to at most a few hundred sensors. 
To improve the scalability of  the SDP method in \cite{Biswas2006}, an alternative (and weaker) second-order cone programming relaxation was proposed in \cite{Tseng2007} that can handle a few thousand sensors. 
%The author, however, reported a run time of about $3$ hours for $2000$ nodes on a standard PC.
A more efficient enhancement of the SDP method that could solve for a few thousand sensors on a standard PC was later proposed in \cite{Wang2008}. 

%Related to this but along a different direction, there is also a large body of literature on the use of distributed (parallel) optimization schemes for directly tackling large-scale SNL problems; 
%e.g., see \cite{Leus2013,Simonetto2012} and the references therein.
%We refer the readers to these recent work \cite{Leus2013,Simonetto2012} and the references therein. 

Our approach in this paper is along the lines of the divide-and-conquer algorithms for anchor-free localization that was proposed in \cite{Koren2005,Zhang2010,Cucuringu2012}. 
In particular, we solve the SNL problem in three steps: (1) we divide the network into overlapping patches (subnetworks), (2) we localize these small patches in parallel 
using some accurate SDP algorithm, and (3) we register the localized patches to determine the sensor positions in a globally-consistent fashion. 
The contribution of this paper concerns step (3). In Section 2, we demonstrate how the non-convex problem of rigid registration (particularly with anchor information) can be relaxed into a tractable convex program. This can be seen as a multi-patch extension of the registration algorithm in \cite{arun1987svd}.
% namely a SDP algorithm for anchor-free registration that was recently proposed in \cite{Chaudhury2013} can be adapted to include anchor constraints, 
% and how this could be applied to the SNL problem in particular. 
%The anchor information is of course used in step (2). 
%We also show that the final localization accuracy can be improved  by also enforcing the anchor constraints during the registration. 
Using the proposed SDP relaxation and performing the subnetwork localizations in parallel, we can solve for 
networks with up to $5000$ sensors within $10$ minutes on a standard PC and a large $8000$-sensor network within $30$ minutes.
Some preliminary simulation results in this direction are provided in Section 3.

% \textbf{Organization}. Next, we present  a least-squares formulation of the registration problem with the anchor constraints, and demonstrate how this non-convex program can be 
% relaxed into an SDP. Simulation results for SNL based on the proposed registration algorithm are presented in Section \ref{secIII}. We conclude the paper with a discussion  in Section \ref{secIV}.

\section{Method}
\label{secII}

We now formally define the SNL problem with anchors, which  includes anchor-free SNL as a special case. 
Suppose we have $N$ sensors and $K$ anchors. Denote the positions of the sensors and anchors by
\begin{equation*}
\x_1,\ldots,\x_N \in \mathbb{R}^d \quad \text{and} \quad  \a_1,\ldots,\a_K \in \mathbb{R}^d,
\end{equation*}
where, usually, $d=2$ or $3$. We will generally refer to the sensors and anchors as \textit{nodes} of the network. 
We are provided the distances between  pairs of nodes that
are within a certain radio range of each other \cite{Biswas2006}.
To represent this distance information, we introduce the \textit{distance graph} $\G=(\V,\E)$ where the vertices $\V=\{1,\ldots,N+K\}$ represent the nodes. 
The first $N$ vertices represent the sensors, while the last $K$ vertices the anchors. 
The edge set $\E \subset \V \times \V$ is given by the requirement that $(k,l) \in \E$ if and only if the distance $d(k,l)$ between nodes $k,l \in \V$ is known. 
Further, we write $\E$ as the union of $\E_{ss}$ and $\E_{sa}$, where $\E_{ss}$ are the sensor-sensor edges and $\E_{sa}$ are the sensor-anchor edges. 
The known distances are represented by $\D = \{d(k,l) : (k,l) \in \E \}$. 
Given $\G, \D,$ and $\A=\{\a_1,\ldots,\a_K\}$, the SNL problem is to determine $\x_1,\ldots,\x_N$ such that 
\begin{equation}
\label{dist_const}
\left.
\begin{aligned}
   \lVert \x_k -\x_l \rVert &= d(k,l), \quad (k,l) \in \E_{ss}  \quad\\ % use \quad as spacer between equation and right brace
   \lVert \x_k -\a_l \rVert &= d(k,l), \quad (k,l) \in \E_{sa}
\end{aligned}
\right\}
\end{equation}
%The constraints \eqref{dist_const} are non-convex in the sensor positions, and this makes the SNL problem difficult. In particular, even determining the feasibility of \eqref{dist_const} is known to be intractable \cite{Saxe79}.   
%In practical settings, however, the distance measurements are usually noisy, and the goal would be to satisfy \eqref{dist_const} approximately \cite{Biswas2006,Leus2013}. 
Motivated by previous work \cite{Koren2005,Zhang2010,Cucuringu2012}, we propose the following divide-and-conquer algorithm.

\subsection{Clustering}

 First, we divide $\V$ into $M$ disjoint clusters by recursively partitioning $\G$ using the Shi-Malik spectral clustering \cite{Shi2000}. 
Of course, other alternative ways of partitioning could be considered. For each cluster, we collect the  
neighbors of every vertex in that cluster ($k$ and $l$ neighbors in $\G$ if $(k,l) \in \E$), and merge a subset of these neighbors  
with the cluster ensuring the size of the augmented cluster to be within a fixed bound (details provided in Section \ref{secIII}). 
More specifically, we pick those neighbors that have the maximum number of edges incident on the cluster, that is, the neighbors that are ``most rigidly'' tied to the cluster. 
The general idea is to expand each cluster so that sufficient pairs of clusters have nodes in common. 
In particular, we now have overlapping \textit{patches} $\P_1,\ldots,\P_M$, where $\P_i \subset \V$. 
By construction, each node belongs to at least one patch, while some nodes belong to two or more patches.  
The latter nodes help in ``propagating'' information between patches during the final registration. 
 
\subsection{Localization} 

We have reduced the large SNL problem into $M$ smaller localization subproblems, one for each patch. 
This is where we speedup the computation by solving these subproblems in parallel. 
More precisely, for $1 \leq i \leq M$, let $\G_i$ be the subgraph of $\G$ induced by the vertices in $\P_i$, $\D_i$  
the distances in $\D$ induced by the edges in $\G_i$, and $\A_i$ the positions of the anchors in patch $\P_i$.  
We use either \texttt{SNLSDP} \cite{Biswas2006} or \texttt{ESDP} \cite{Wang2008} (which is a further relaxation of \texttt{SNLSDP}) 
to compute the positions of the sensor vertices in $\P_i$ from the knowledge of $\G_i,\D_i,$ and $\A_i$. 
For large noise and small sensing radius, \texttt{SNLSDP} occasionally fails to localize certain patches (the corresponding SDP is infeasible). 
We localize these exceptional patches using  \texttt{ESDP}, which is somewhat less accurate than \texttt{SNLSDP} but has a larger scope.
While one can also use other efficient SNL algorithms, these SDP-based solvers suit our purpose as they offer a good tradeoff between accuracy and run-time 
for small problems. Finally, we refine the sensor positions  using the local optimization in \cite{Biswas2006}.

At the end of this phase, all the patches have been positioned in independent coordinate systems. 
If the $k$-th sensor belongs to patch $\P_i$, then we use $\x_{k,i}$ to denote the position of that sensor in $\P_i$.  
In the ideal setting where each patch graph $\G_i$ is uniquely localizable \cite{So2007} and the distances are noise-free, the local sensor 
positions $\{\x_{k,i}: k \in \P_i\}$ are identical to the global positions $\{\x_k: k \in \P_i\}$ up to a rigid transform that fixes the anchors \cite{Biswas2006,So2007}. That is, for some 
orthogonal transform $\O_i$ and translation $\t_i$,  
\begin{equation}
\label{xki}
\x_k=  \O_i \x_{k,i}  + \t_i \qquad (k \in \P_i),
\end{equation}
and
\begin{equation}
\label{FP}
\a_l =  \O_i \a_l + \t_i \qquad (l \in \P_i).
\end{equation}
It is understood here and henceforth that $k \in \{1,\ldots,N\}$ and $l \in \{N+1, \ldots ,N+K\}$. 
%Note that $\O_i$ and $\t_i$ can be arbitrary if $\P_i$ has no anchors, as is the case with anchor-free 
%localization. At the other extreme, if $\P_i$ has $d+1$ or more (non-degenerate) anchors, then $\O_i$ must be identity and $\t_i=\mathbf{0}$. 

\subsection{Rigid Registration}

It remains to determine $\x_1,\ldots,\x_N$ from the system of equations in \eqref{xki} and \eqref{FP}. In practice, 
%there could be noise in the distance measurements, or some of 
%the patch graphs  arising from the clustering could fail to be uniquely localizable. Even the positions of the anchors that are determined using GPS are prone to error due to the limited GPS accuracy.
%In such cases, 
one would expect these equations to hold only approximately due to various imperfections. Thus, one would like to have a reconstruction in which the discrepancy 
from \eqref{xki} and \eqref{FP} is as small as possible.  
In particular, we consider the following quadratic loss $\phi$ given by
\begin{eqnarray*}
\begin{aligned}
\phi=\sum_{i=1}^M \Big\{   \sum_{k \in \P_i} \ \lVert \x_k - &\O_i \x_k^{(i)} - \t_i \rVert^2   \\
&+  \lambda  \sum_{l \in \P_i} \ \lVert \O_{M+1} \a_l -  \O_i \a_l - \t_i \rVert^2 \Big\},
\end{aligned}
\end{eqnarray*}
where the optimization variables are: $\x_1,\ldots,\x_N; \t_1,\ldots,\t_M \in \mathbb{R}^d$ and $\O_1,\ldots,\O_{M+1} \in \mathbb{O}(d)$.
Here and henceforth $\mathbb{O}(d)$ denotes the group of $d \times d$ orthogonal matrices, $\lVert \cdot \rVert$ is the Euclidean norm, and $\lambda >0$  is used to balance the loss. The slack variable $\O_{M+1}$  is introduced to make $\phi$ homogeneous with respect to the variables. 

The above optimization can be seen as a generalization of the two-patch registration optimization  addressed in \cite{arun1987svd}. In fact, the present idea of first optimizing over the translations and then over the orthogonal transforms is similar to the strategy used in this paper.
%We note that a similar homogenization was used for the MIMO detection problem in \cite{Zhang2010review}.
We first massage $\phi$ into something more compact using matrix notations.
We begin by collecting the free variables (sensor positions and translations) and the constrained variables (orthogonal transforms) into two separate matrix variables:
\begin{eqnarray}
\label{defZ}
 \Z &=& \big[ \x_1 \cdots \x_N \ \ \t_1 \cdots \t_M  \big] \in \mathbb{R}^{d \times (N+M)}, \\ \nonumber
 \O &= &\big[\O_1 \cdots \O_{M+1}]  \in \mathbb{R}^{d \times (M+1)d}.
\end{eqnarray}
Next, we introduce an undirected bipartite graph $\G = (\V_x \cup \V_P, \E)$, where the vertices $\V_x = [1,N+K]$ correspond to the 
sensor and anchor, and the vertices $\V_P = [1,M]$ correspond to the patches. The edge set $E \subset \V_x \cup \V_P$ is given by the requirement that $(k,i) \in \E$ if and only if the $k$-th node is in patch $\P_i$. To distinguish between the sensors and anchors, we further divide $\V_x$ into the sensor vertices $\V_s=[1,N]$ and the anchor vertices $\V_a=[N+1,N+K]$. We denote the number of anchors in patch $\P_i$ by $K_i$.

Let $\Del^{L}_i$ denote the all-zero vector of length $L$ with $1$ at the $i$-th coordinate, and define
\begin{equation*}
 \e_{ki} = \Del^{N+M}_k -  \Del^{N+M}_{N+i}  \quad \text{and} \quad \f_j=\Del^{M+1}_{M+1}-\Del^{M+1}_j .
\end{equation*}
In terms of the above notations, we can then write
\begin{eqnarray*}
\phi(\Z,\O) &= \sum_{k \in \V_s} \sum_{(k,i) \in \E} \ \lVert \Z \e_{ki} - \O (\Del^{M+1}_i \otimes \I_d) \x_{k,i} \rVert^2  \\
&+    \sum_{l \in \V_a} \sum_{(l,j) \in \E} \ \lVert \Z \Del^{N+M}_{N+j} - \O (\f_j \otimes \I_d) \a_l   \rVert^2,
\end{eqnarray*}
where $\I_d$ is the $d \times d$ identity matrix and $\otimes$ is the Kronecker product. By expanding out the squares and rearranging the terms, we get after some computation
\begin{equation}
\label{phiZO}
\phi(\Z,\O) = \Tr \left(\begin{bmatrix} \Z& \O  \end{bmatrix} \begin{bmatrix} \J & -\B^T \\ -\B & \mathbf{D} \end{bmatrix} \begin{bmatrix} \Z^T\\ \O^T \end{bmatrix} \right)
\end{equation}
where
\begin{eqnarray*}
\label{coeff}
\begin{aligned}
\J &= \sum_{k \in \V_s} \sum_{(k,i) \in \E} \e_{k,i} \e_{k,i}^T +   \sum_{l \in \V_a} \sum_{(l,j) \in \E} \Del^{N+M}_{N+j} {\Del^{N+M}_{N+j}}^T, \\ 
\B &= \sum_{k \in \V_s} \sum_{(k,i) \in \E} (\Del^{M+1}_i \otimes \I_d) \x_{k,i} \e_{k,i}^T  \\ 
&+   \sum_{l \in \V_a} \sum_{(l,j) \in \E} (\f_j \otimes \I_d) \a_l {\Del^{N+M}_{N+j}}^T, \\ 
\mathbf{D} &= \sum_{k \in V_s} \sum_{(k,i) \in \E} (\Del^{M+1}_i \otimes \I_d) \x_{k,i} \x_{k,i}^T (\Del^{M+1}_i \otimes \I_d)^T  \\
&+   \sum_{l \in \V_a} \sum_{(l,j) \in \E} (\f_j  \otimes \I_d) \a_l \a_l^T (\f_j  \otimes \I_d)^T.
\end{aligned}
\end{eqnarray*}
Note that the block matrix $\J$ is of size $(N+M) \times (N+M)$, $\B$ is of size $(M+1)d \times (N+M)$, and $\mathbf{D}$ is of size $(M+1)d \times (M+1)d$. In fact, $\J = \L +    \mathrm{diag}(0,\ldots,0,K_1,\ldots,K_M)$, where $\L$ is the Laplacian of the bipartite graph $\G'$ that is obtained by removing the anchor vertices $\V_a$ and the corresponding edges from $\G$. 
In particular, $\J$ is positive semidefinite. Moreover, if $\G'$ is connected and some $K_i>0$, then $\J$ is non-singular.

The optimization problem can now be compactly expressed as
\begin{equation*}
\min \ \Big\{ \phi(\Z, \O) : \Z \in \mathbb{R}^{d \times (N+M)}, \ \O \in \mathbb{R}^{d \times (M+1)d}, \  \O_i \in \mathbb{O}(d) \Big\}.
\end{equation*}
where $\O_i$ denotes the $i$-th $d \times d$ block of $\O$. The present strategy is to first solve for the unconstrained variable $\Z$ in terms of the unknown orthogonal transformations $\O$, representing the former as linear combinations of the latter.
In particular, fix some arbitrary $\O$, and let $\psi(\Z)=\phi(\Z,\O)$. It is clear from \eqref{phiZO} that $\psi(\Z)$ is quadratic in $\Z$. In particular, the stationary points 
$\Z^{\star} = \Z^{\star}(\O)$ of $\psi(\Z)$ obtained by setting its gradient to zero are given by $\Z^{\star} \J=\O \B$.
Note that the Hessian of $\psi(\Z)$ equals $2\J$, and we known that $\J$ is non-singular if $\G'$ is connected (this is always true in practice) and if there is at least one anchor. 
Therefore, in this case the unique minimizer of $\psi(\Z)$ is given by
\begin{equation}
\label{inversion}
\Z^{\star} = \O \B \J^{-1}.
\end{equation}
Substituting $\Z^{\star} $ in \eqref{phiZO} and  simplifying, we have $\psi(\Z^{\star}) = \phi(\Z^{\star},\O)= \Tr(\C \O^T \O)$,
where $\C  = \mathbf{D} - \B \J^{-1} \B^T$. In other words, denoting the $(i,j)$-th block of $\C$ by $\C_{ij}$, we have thus reduced the original problem to that of minimizing
\begin{equation}
\label{optO}
\Tr(\C \O^T \O)=\sum_{i=1}^{M+1} \sum_{j=1}^{M+1} \Tr( \C_{ij} \O_i^T \O_j)
\end{equation}
where the variables are the orthogonal matrices $\O_1,\ldots,\O_{M+1}$. This is clearly a difficult non-convex problem. We now present a convex relaxation of this problem following the ideas  in \cite{Chaudhury2013}.
In particular, we consider the (block) Gram matrix of the transforms, namely $\mathbf{G}=\O^T\O$, which is of size ${(M+1)d \times (M+1)d}$ and whose $(i,j)$-th block is given by
$\mathbf{G}_{i,j} = \O_i^T \O_j$. In terms of this Gram matrix, we can equivalently formulate the optimization in \eqref{optO} as
\begin{eqnarray*}
\begin{aligned}
& \min  \ \ && \Tr(\C\mathbf{G}) \\
&  \text{subject to} 
&& \mathbf{G} \succeq 0, \mathrm{rank}(\mathbf{G})=d, \mathbf{G}_{ii} = \I_{d} \ (i=1,\ldots,M+1).
\end{aligned}
\end{eqnarray*}
By dropping the non-convex rank constraint, we get the the following convex program:
\begin{eqnarray}
\label{manopt}
\begin{aligned}
& \min  \ \ && \Tr(\C\mathbf{G})   \\
& \text{subject to}
& & \mathbf{G} \succeq 0 , \ \mathbf{G}_{ii} = \I_{d} \ (i=1,\ldots,M+1).
\end{aligned}
\end{eqnarray}
Here $\mathbf{G} \succeq 0 $ means that $\mathbf{G}$ is symmetric and positive semidefinite, and $\mathbf{G}_{ii} = \I_{d}$ forces the diagonal $d \times d$ blocks of $\mathbf{G}$ to be identity matrices (thus requiring each transform to be orthogonal).
Now \eqref{manopt} is a standard convex program called a semidefinite program that has been well-studied \cite{Boyd1996}. 
Suppose that $\mathbf{G}^{\star}$ is the global minimizer of \eqref{manopt}.
By the diagonal block constraints in \eqref{manopt}, it follows that $\mathrm{rank}(\Gs) \geq d$. If $\mathrm{rank}(\Gs)$ is exactly $d$, we have in fact solved the original non-convex problem (relaxation is tight). In particular, the factorization $\Gs = {\O^{\star}}^T \O^{\star}$, where $\O^{\star}$ has rank $d$, gives us the desired orthogonal transforms. 
Following \eqref{defZ} and \eqref{inversion}, we set $\Z^{\star} = \O^{\star} \B \J^{-1}$ and take the first $N$ columns of $\Z^{\star} $ are taken to be the reconstructed global coordinates.

 \begin{figure}
\centering
\subfloat[$50$ anchors, before refinement.]{\includegraphics[width=0.5\linewidth]{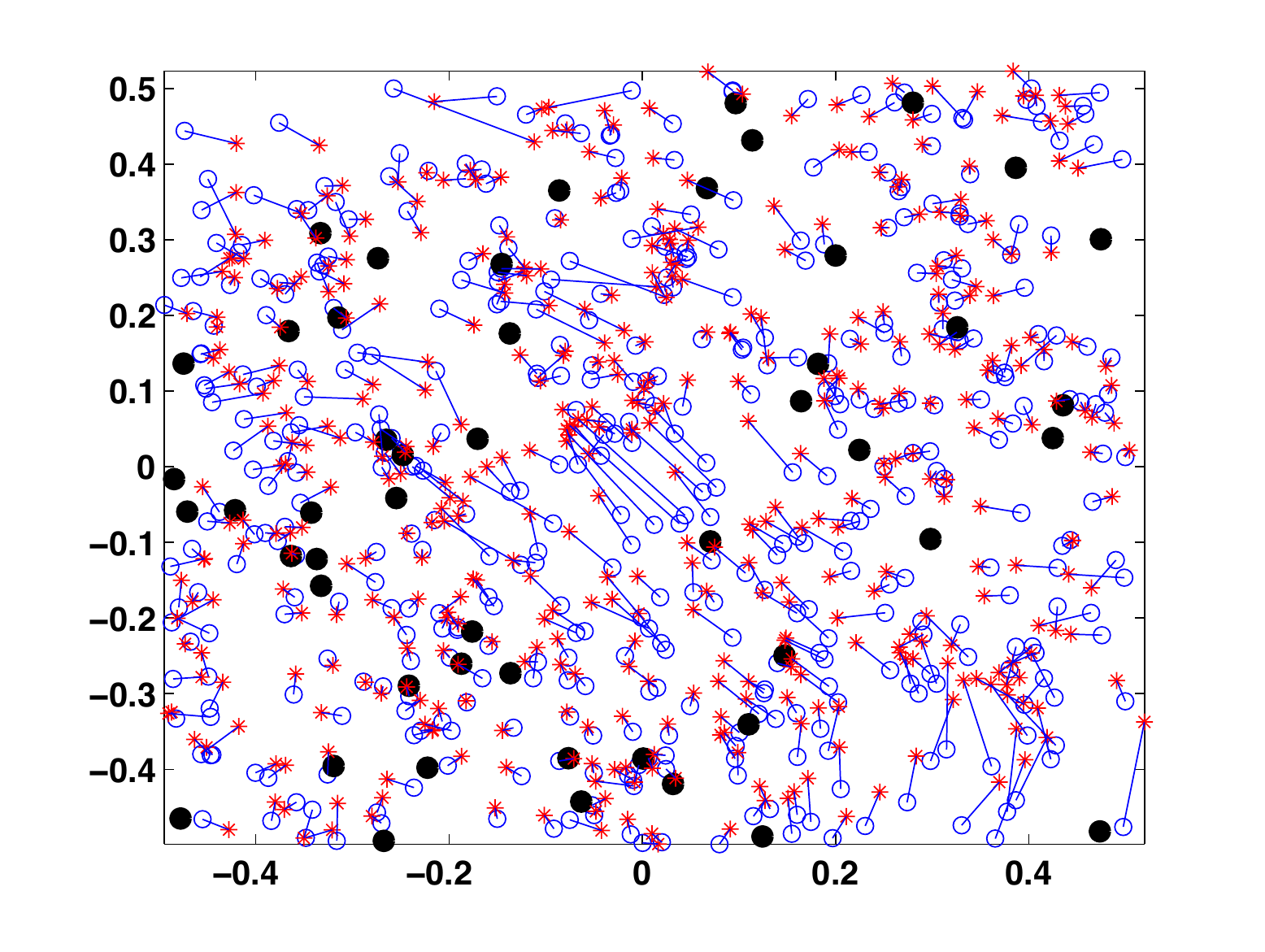}} 
\subfloat[$50$ anchors, after refinement.]{\includegraphics[width=0.5\linewidth]{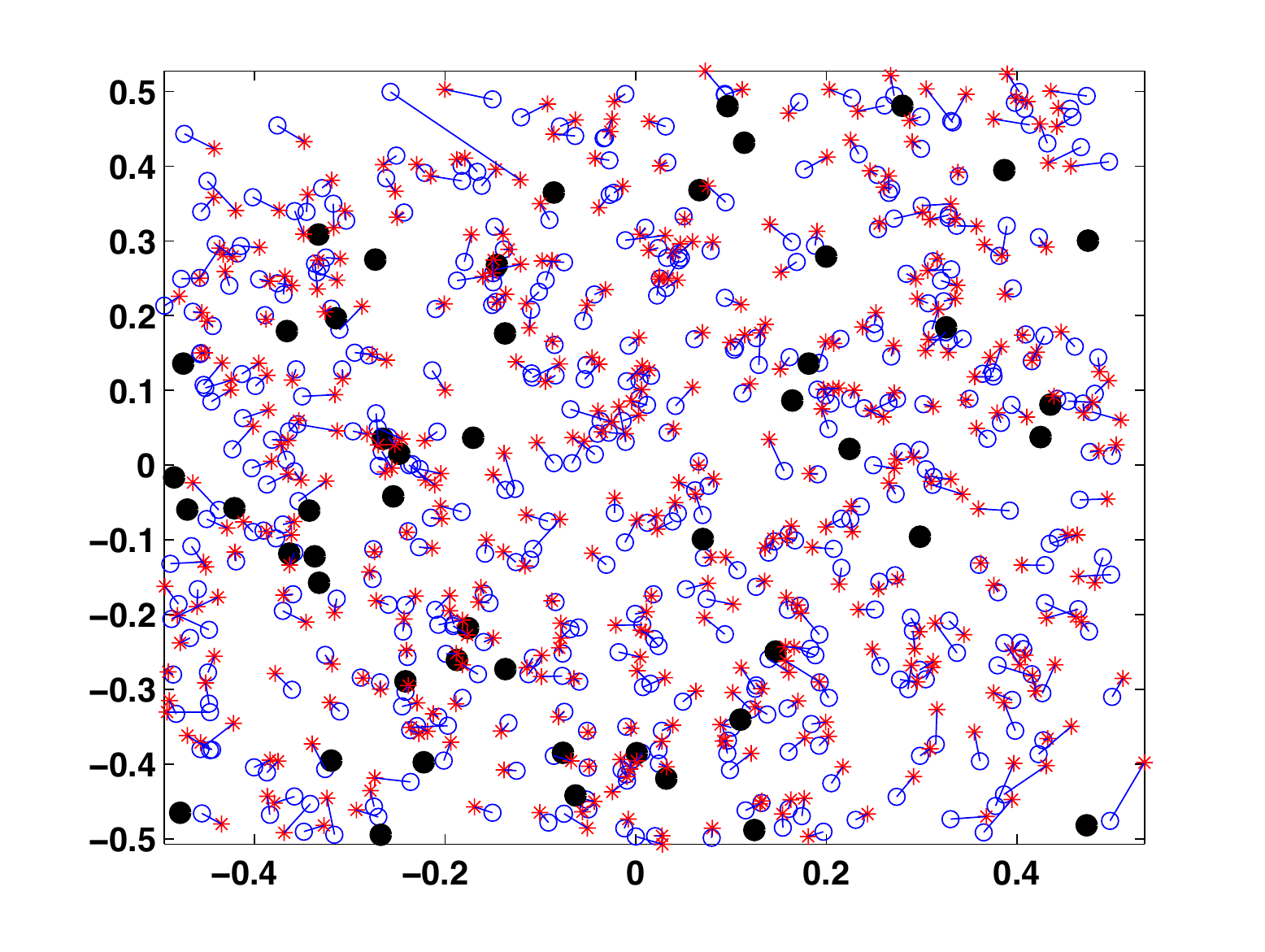}} \\ \vspace{-0.3cm}
\subfloat[$3$ anchors, before refinement.]{\includegraphics[width=0.5\linewidth]{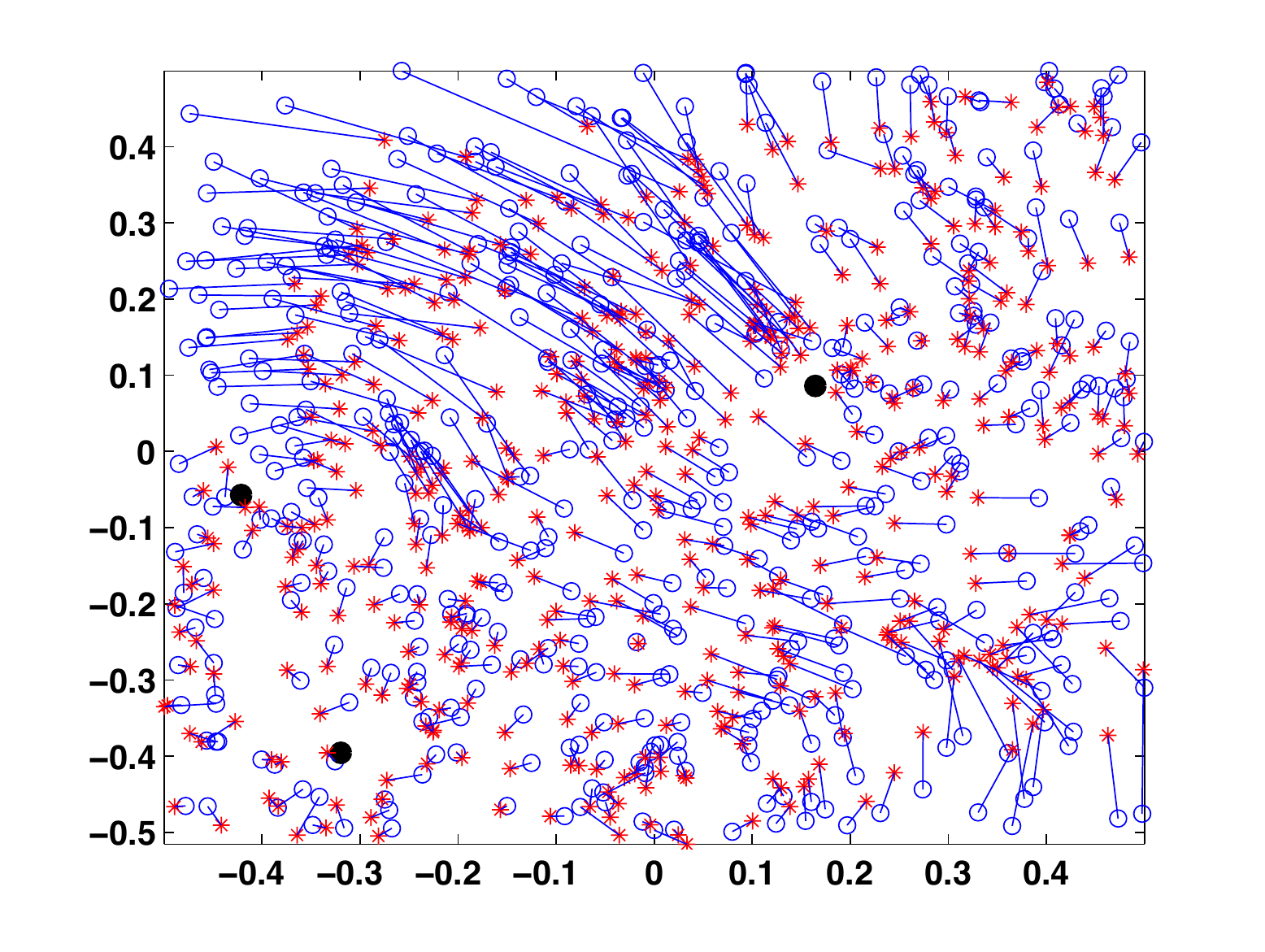}}
\subfloat[$3$ anchors, after refinement.]{\includegraphics[width=0.5\linewidth]{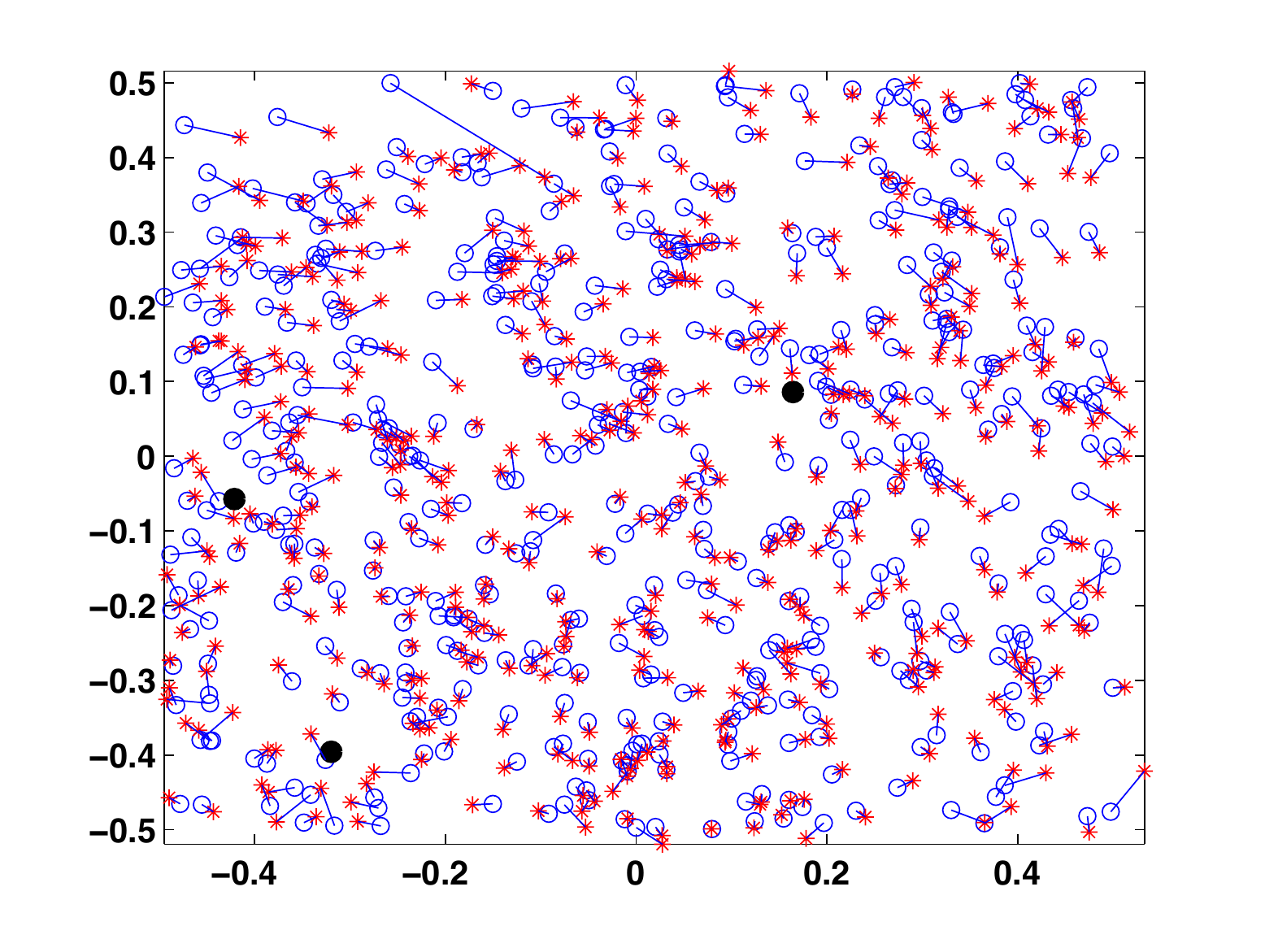}} 
\caption{Localization results for \texttt{SNLSR} using full and minimal anchor information during registration. We used a minimum of $3$ anchors (randomly picked from the full set of anchors) to fix the global rigid transform. 
The problem parameters are: $N=500, r=0.2$, and $\eta=0.5$. 
The RMSD's are: (a) \textbf{3.5e-2}, (b) \textbf{2.3e-2}, (c) \textbf{7.2e-2}, and (d) \textbf{2.5e-2}.
The true sensor positions are marked with blue circles, the anchors with solid black circles, and the estimated sensors with red stars. 
The true and estimated sensors are joined by solid blue lines.}
\label{Figure1}
\end{figure}

On the other hand, if $\mathrm{rank}(\Gs) > d$, we need to project $\Gs$ onto the space of Gram matrices of orthogonal transforms. 
In particular, let $\lambda_1 \geq \lambda_2 \geq \cdots  \geq \lambda_{(M+1)d} \geq 0$ be the eigenvalues of $\Gs$, and $\q_1,\ldots,\q_{(M+1)d}$ be the corresponding eigenvectors. Let
\begin{equation*}
\W = \big[ \sqrt{\lambda_1} \ \q_1 \  \cdots \ \sqrt{\lambda_d} \  \q_d \big]^T \in \mathbb{R}^{d \times (M+1)d}.
\end{equation*}
Notice that due to the relaxation, the $d \times d$ blocks of $\W$ are not guaranteed to be orthogonal. We round each $d \times d$ block of $\W$ to its ``closest'' orthogonal matrix. More precisely, let $\W  = [\W_1 \cdots \W_{M+1}]$.
For every $i=1,\ldots,M+1$, we find $\O_i^{\star} \in \mathbb{O}(d)$ that minimizes $\lVert   \O_i - \W _i \rVert_F$ where $\lVert \cdot \rVert_F$ denotes the Frobenius norm.
This has a closed-form solution, namely $\O_i^{\star}=\mathbf{U} \mathbf{V}^T,$ where $\mathbf{U} \mathbf{\Sigma} \mathbf{V}^T$ is the SVD of $\W _i$ \cite{arun1987svd}. Following \eqref{defZ} and \eqref{inversion}, we form the matrix $\O^{\star} = \big[ \O^{\star}_1 \ldots \O^{\star}_{M+1} \big]$
and take the first $N$ columns of $\Z^{\star} = \O^{\star} \B \J^{-1}$  to be the sensor coordinates.

In the final step, we refine the sensor positions obtained at the end of registration using the gradient-based local search in \cite{Biswas2006}. 
As we will see, this step is in fact quite effective in improving the localization accuracy. 
We denote the final sensor positions by  $\hat{\x}_1,\ldots,\hat{\x}_N$. 

Henceforth, we will refer to the proposed algorithm as \texttt{SNLSR}, short for ``SNL via Subnetwork Registration''. 

\section{Numerical Simulations}
\label{secIII}

We now present some simulation results for SNL in $\mathbb{R}^2$ using \texttt{SNLSR}. 
All the simulations were carried out in Matlab $8.1$ on a four-core $2.83$ GHz Linux workstation with a $3.6$ GB memory. 
It is clear that the computation-intensive steps of our approach are the determination of the patch localizations and the solution of \eqref{manopt}. 
The former was accomplished in a parallel fashion using the Matlab implementations of \texttt{SNLSDP} \cite{Biswas2006} and \texttt{ESDP} \cite{Wang2008}.

For solving the SDP relaxation of \eqref{manopt}, we used the interior-point solver \texttt{SeDuMi} \cite{SeDuMi} for medium-sized problems, and \texttt{SDPLR} \cite{SDPLR} for large problems.
For the hierarchical clustering, we partitioned $\G$ in a recursive fashion until the size of each cluster is below $30$. 
We next grow each cluster into a patch by adding its neighbors as explained in Section \ref{secII}, where we limit the patch size to $45$. 
This is roughly the largest patch size for which the run time of \texttt{SNLSDP} is reasonable.
%Due to this bound on patch size, the number of patches in our implementation scale almost linearly with the number of sensors and anchors.
%It is in principle possible to use bigger patches and reduce the number of patches, e.g., when $N>2000$, but we have not investigated this possibility.    

\begin{figure}
\centering
\subfloat[$\eta=0$,   \textbf{1.1e-4} (\texttt{SNLSR}).]{\includegraphics[width=0.5\linewidth]{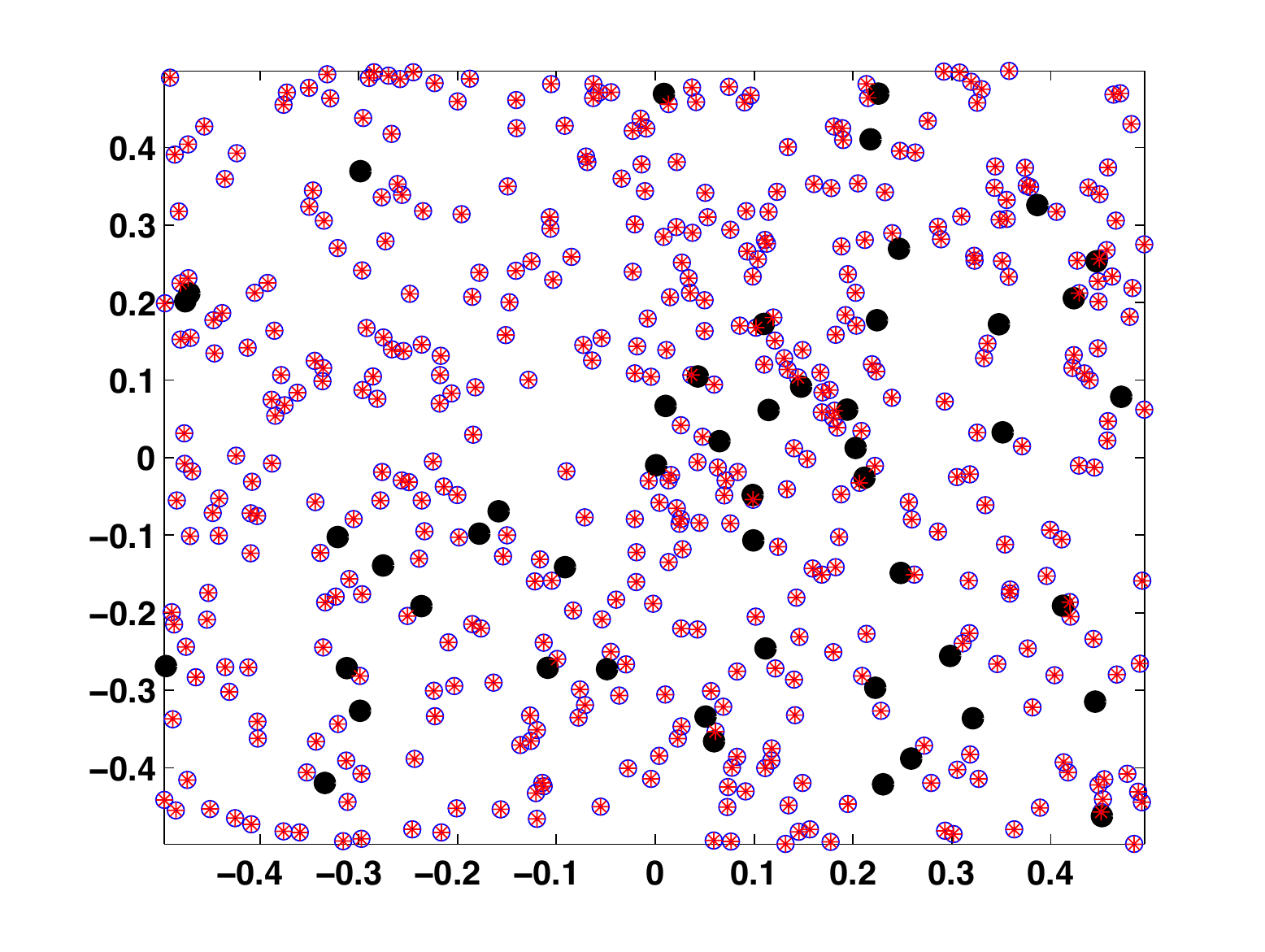}} 
\subfloat[$\eta=0$,   \textbf{5.4e-7} (\texttt{ESDP}).]{\includegraphics[width=0.5\linewidth]{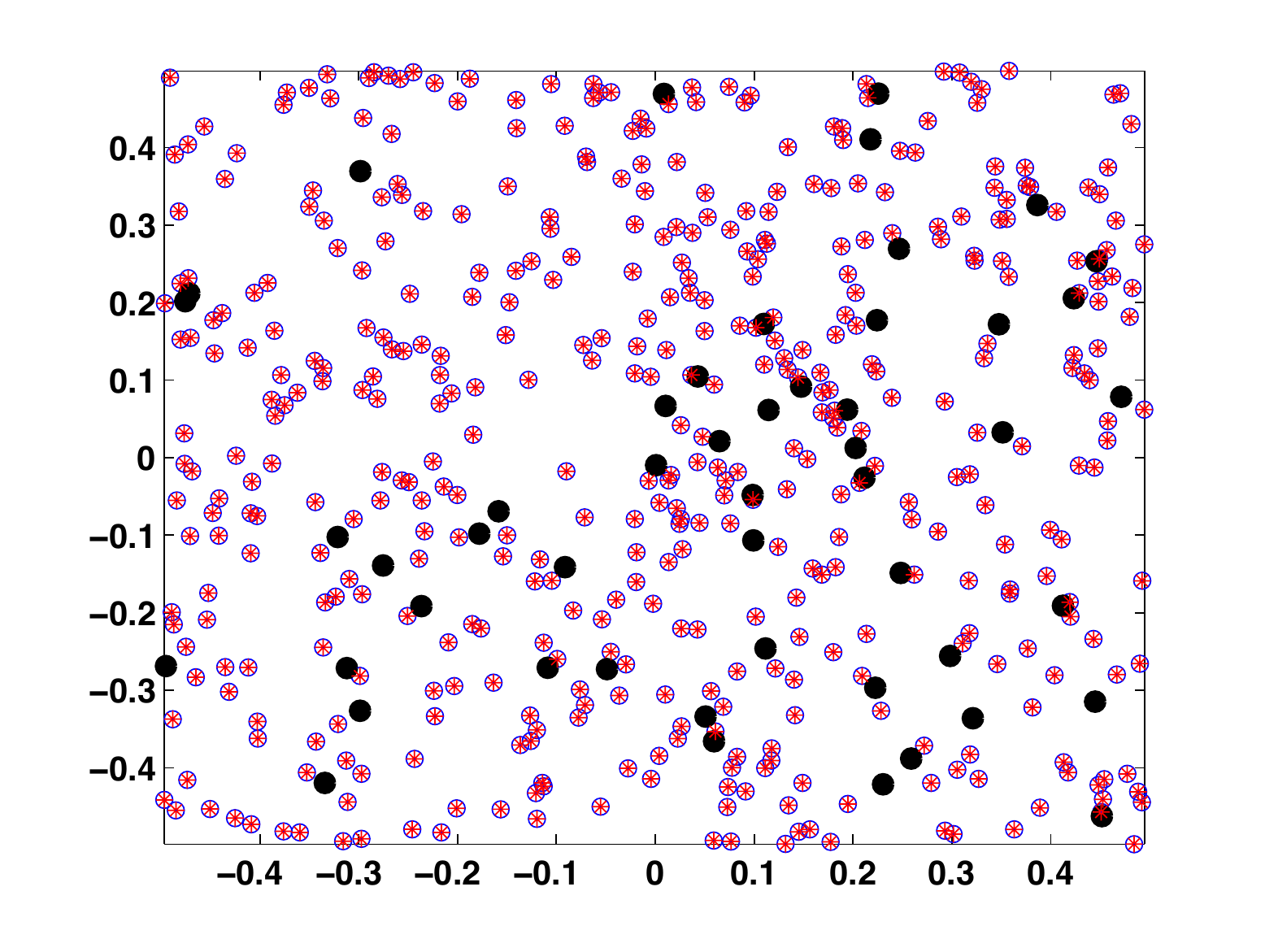}} \\ \vspace{-0.3cm}
\subfloat[$\eta=0.3$, \textbf{1.3e-2} (\texttt{SNLSR}).]{\includegraphics[width=0.5\linewidth]{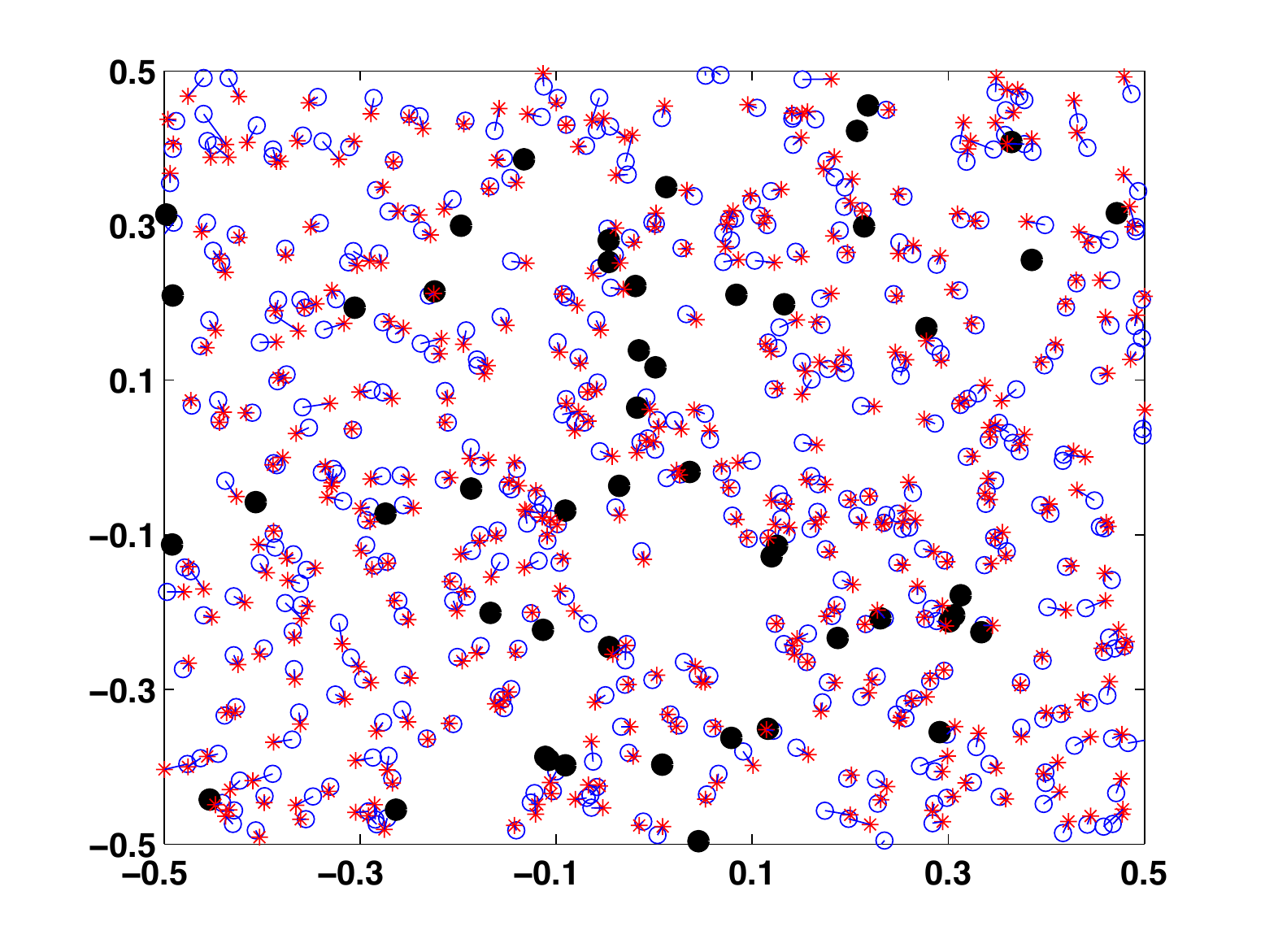}} 
\subfloat[$\eta=0.3$, \textbf{5.5e-2} (\texttt{ESDP}).]{\includegraphics[width=0.5\linewidth]{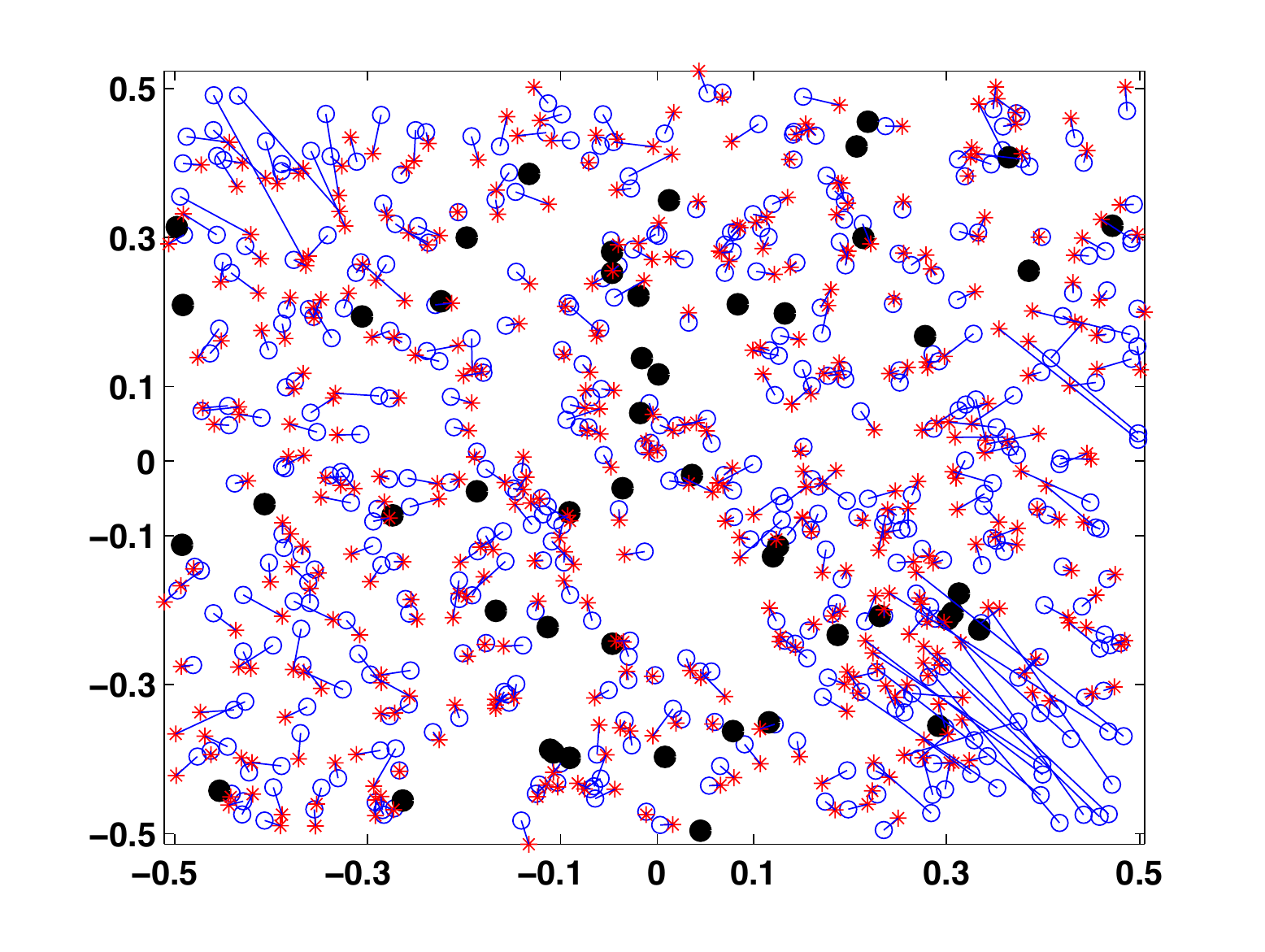}}
\caption{Localization using \texttt{SNLSR} and \texttt{ESDP} \cite{Wang2008}. The RMSD's are shown in bold. The problem parameters in either case are: $N=500, K= 50$, and $r=0.2$ (around $10\%$ edges).  
The overall run time of \texttt{SNLSR} was about 10 and 18 seconds for the clean and noisy problems; the corresponding times for \texttt{ESDP} were about 2 minutes and 38 seconds.
%(the fact that \texttt{ESDP} takes less time
%for noisy problems was also reported in \cite{Wang2008}). 
%The results shown here are those obtained after refining the localizations of either method using gradient descent.
} 
\label{Figure2}
\end{figure}

Following \cite{Biswas2006,Wang2008}, we generate the true positions of the sensors and anchors by drawing $|\V| = N+K$ points $\{\x_i : i \in \V \}$ from the uniform distribution on the unit square $[-0.5,0.5]^2$. 
We take the first $N$ points to be the sensors and the remaining $K$  points to be the anchors. Following \cite{Tseng2007}, we set $K=[N/10]$.
The distance graph $\G=(\V,\E)$ is given by the condition that $(k,l) \in \E$ if and only if $\lVert \x_k - \x_l \rVert \leq r$, where $r \in (0,1)$ is the radio range. 
We also use the noise model from \cite{Biswas2006,Wang2008} in which the measured distances are given by $ d(k,l) = |1 + \eta \cdot N(0,1) | \cdot \lVert \x_k - \x_l \rVert$ for $(k,l) \in \E,$
% \begin{equation*}
% d(k,l) = |1 + \eta \cdot N(0,1) | \cdot \lVert \x_k - \x_l \rVert \qquad (k,l) \in \E, 
% \end{equation*}
where $N(0,1)$ is the standard normal distribution and $\eta \in (0,1)$ is the noise level. While other noise models could also be considered,  we settled for this noise model to 
facilitate comparison with the results in \cite{Biswas2006,Wang2008}. For the same reason, we use the Root-Mean-Squared-Deviation (RMSD) to quantify the localization accuracy which is
given by the square root of $(1/N)\sum_{k=1}^N \lVert \hat{\x}_k -\x_k  \rVert^2$. 

Note that it is possible to register the patches in a globally consistent manner using the anchor-free registration, simply by dropping the equations in \eqref{FP}. 
Do we get better localization accuracy by incorporating the anchor information into the registration? 
Exhaustive simulations (not reported here) show that by incorporating anchors into the registration, it is indeed possible to improve the accuracy. 
%Of course, the margin of improvement is determined by the noise level and the graph connectivity. 
Simulations show that if the noise is small and the patches are sufficiently rigid, then the margin of improvement is small.
However, under more adversarial settings, the margin can be quite substantial. 
A particular example is presented in Figure \ref{Figure1}. 
Notice the poor localization in (c) around the top left corner. This is because a couple of patches around this region 
were poorly localized by \texttt{SNLSDP}, and moreover, the anchors in this region were not considered during the registration. 
As a consequence, the rigid transform associated with these patches were poorly estimated during the registration.
Also, notice that the gradient-based refinement is quite effective in reducing the RMSD in either case. In particular, while the RMSD gap is about $50\%$ after the registration, 
the gap comes down to about $20\%$ after the refinement.

\begin{table}
\caption{Comparison of the run time and localization accuracy of \texttt{SNLSR} ($t1$ and RMSD1) with \cite{Biswas2006,Wang2008} ($t2$ and RMSD2) for the
unit-square graph. 
We used \texttt{SNLSDP} \cite{Biswas2006} for $N < 150$, and \texttt{ESDP} \cite{Wang2008} for larger problems (see text for further details). 
%The simulations were performed using Matlab 8.1 on a four-core Linux workstation ($2.83$ GHz and $3.6$ GB memory). 
%The results were averaged over $10$ realizations.
} 
\vspace{1mm}
\centering 
\begin{tabular}{| l | l | l | l | l | l | l |}
\hline
    
    $N$   & $r$   & $\eta$ &  $t1$ & $t2$  & RMSD1  & RMSD2 \\ \hline \hline

    100   & 0.4     & 0    &  8.1 sec    &  9.1 sec        &  8.9e-8  &  2.7e-9 \\ \hline
    
    100   & 0.4     & 0.3  &  4.2 sec        &  8.6 sec       &  3.1e-2  &  3.2e-2 \\ \hline
    
    150   & 0.3     & 0    &   5.1 sec        &  38.9 sec       &  3.8e-8  &   1.4e-8 \\ \hline
    
    150   & 0.3     & 0.2  &  5.6 sec       &  91.3 sec         &  2.1e-2  &  1.9e-2 \\ \hline
    
    150   & 0.2     & 0.1  &  6.8 sec        &  2.1 min              &  2.5e-2  &  2.1e-1 \\ \hline
    
    500   & 0.2   & 0      &  10.4 sec   &  2.74 min     &  3.8e-6 &  1.2e-7 \\ \hline
    
    500   & 0.2   & 0.01    &  21.5 sec  &  46.1   sec   &  4.3e-4  &  8.2e-4 \\ \hline
    
    500   & 0.2   & 0.1    &  15.1 sec  &  37.9 sec      &  4.3e-3  &  2.3e-2 \\ \hline
        
    1000  & 0.06   & 0    &  32.7 sec    &  9.35 min    &  1.9e-2 &  1.3e-2 \\ \hline
    
    1000  & 0.06   & 0.01    &  33.1 sec &  5.55 min      &  3.4e-2 &  3.2e-2 \\ \hline
    
    1000  & 0.06   & 0.05    &  25.7 sec   &  1.82 min     &  2.8e-2 &  1.3e-2 \\ \hline
    
    2000  & 0.04      & 0    & 1.84 min    &  11.35 min    &  1.3e-2  &  1.2e-2 \\ \hline
    
   2000  & 0.04   & 0.005    &   1.93 min  &  6.54 min      &  1.8e-2 &  1.1e-2 \\ \hline
    
    2000  & 0.04   & 0.05    &  1.62 min   &  7.26 min   &  1.5e-2 &  1.1e-2 \\ \hline
    
    4000  & 0.03   & 0    &   6.82 min   &  24.35 min    &  1.1e-2  &  1.6e-2\\ \hline
    
 %  4000  & 0.03   & 0.001    & 8.69 min  &  33.76 min      &  1.1e-2 &  6.1e-3 \\ \hline
    
  4000  & 0.03   & 0.01    &  6.81 min  &  22.54 min      &  1.2e-2 &  8.2e-3 \\ \hline
  
  8000  & 0.02  & 0       &  28.34 min   &  1 hr 16 min       &  2.4e-3 &  4.8e-3 \\ \hline
\end{tabular}
\label{table1}
\end{table}

We next compare \texttt{SNLSR} with the SDP-based methods in \cite{Biswas2006,Wang2008}, both in terms of accuracy and scalability. 
Using the computational resources mentioned earlier, we could solve for at most $N=150$ sensors using \texttt{SNLSDP} \cite{Biswas2006}. 
We used \texttt{ESDP} to address larger problems. 
Simulations suggest that our divide-and-conquer method is significantly faster (often by an order) than these methods. 
%These will be investigated in future work.
The accuracy of \texttt{SNLSR} is comparable, and occasionally better, than these methods. A visual comparison for a moderate-sized problem is provided in Figure \ref{Figure2}. 
Further comparisons are provided in Table \ref{table1}. We note that we empirically tuned $\lambda$ in $\phi$ to get the minimum RMSD from \texttt{SNLSR}. 
For the SNL settings considered here, the optimal $\lambda$ was in the interval $(1,4)$, and the variability of the RMSD within this interval was small (within $10\%$ of the optimal), and even smaller
after the refinement.

\section{Conclusion}

We presented a divide-and-conquer approach for SNL and demonstrated its utility for large-scale problems.
In particular, we showed how the non-convex problem of registering the subnetworks can be relaxed and solved efficiently using modern convex programming tools.
While the simulation results presented here are far from exhaustive,
%For example, it is not clear how the proposed method compares with algorithms that are not based on 
%optimization, e.g., those based on trilateration \cite{Patwari2005,Madria2003}. Even for the optimization-based methods, it is yet to be seen how \texttt{SNLSR} compares, for example, with those in \cite{Tseng2007,Montanari2011}. 
they  nevertheless demonstrate that the idea of localizing patches in parallel and then registering them in a globally consistent fashion can indeed lead to fast and scalable algorithms. 

\vfill\pagebreak

\bibliographystyle{IEEEbib}

\end{document}